\documentclass[10pt, conference, compsocconf]{IEEEtran}

\usepackage{algorithm,algorithmic,amsbsy,amsmath,amssymb,epsfig,subfig,bbm,mathrsfs,mathdots,arydshln,arydshln,multirow,verbatim,booktabs,setspace,url,stfloats,cite,nomencl,epstopdf, comment}

\usepackage{braket}

\usepackage{xcolor}
\def\BibTeX{{\rm B\kern-.05em{\sc i\kern-.025em b}\kern-.08em
    T\kern-.1667em\lower.7ex\hbox{E}\kern-.125emX}}

\newcommand{\beq}{\begin{equation}}
\newcommand{\eeq}{\end{equation}}
\newcommand{\bitm}{\begin{itemize}}
\newcommand{\ba}{\begin{array}}
\newcommand{\ea}{\end{array}}
\newcommand{\eitm}{\end{itemize}}
\newcommand{\beqn}{\begin{eqnarray}}
\newcommand{\eeqn}{\end{eqnarray}}
\newcommand{\beqno}{\begin{eqnarray*}}
\newcommand{\eeqno}{\end{eqnarray*}}
\newcommand{\bma}{\begin{displaymath}}
\newcommand{\ema}{\end{displaymath}}
\newcommand{\bnu}{\begin{enumerate}}
\newcommand{\enu}{\end{enumerate}}
\newcommand{\bce}{\begin{center}}
\newcommand{\ece}{\end{center}}
\newcommand{\btb}{\begin{tabular}}
\newcommand{\etb}{\end{tabular}}

\begin{document}

%\title{Entangled Pair Resource Allocation under Uncertain Fidelity Requirements}
\title{Entangled Pair Resource Allocation under Uncertain Fidelity Requirements}
%\begin{comment}
\author{\IEEEauthorblockN{Rakpong Kaewpuang$^{\mathrm{1}}$, Minrui Xu$^{\mathrm{1}}$, Stephen John Turner$^{\mathrm{2}}$, Dusit Niyato$^{\mathrm{1}}$, Han Yu$^{\mathrm{1}}$, and Dong In Kim$^{\mathrm{3}}$ } \\
\IEEEauthorblockA{ $^{\mathrm{1}}$School of Computer Science and Engineering, Nanyang Technological University, Singapore \\
                    $^{\mathrm{2}}$School of Information Science and Technology, Vidyasirimedhi Institute of Science and Technology, Thailand \\
                    $^{\mathrm{3}}$School of Information and Communication Engineering, Sungkyunkwan University, South Korea
       }}
%\end{comment}
\maketitle

\begin{abstract}
In quantum networks, effective entanglement routing facilitates remote entanglement communication between quantum source and quantum destination nodes. Unlike routing in classical networks, entanglement routing in quantum networks must consider the quality of entanglement qubits (i.e., entanglement fidelity), presenting a challenge in ensuring entanglement fidelity over extended distances. To address this issue, we propose a resource allocation model for entangled pairs and an entanglement routing model with a fidelity guarantee. This approach jointly optimizes entangled resources (i.e., entangled pairs) and entanglement routing to support applications in quantum networks. Our proposed model is formulated using two-stage stochastic programming, taking into account the uncertainty of quantum application requirements. Aiming to minimize the total cost, our model ensures efficient utilization of entangled pairs and energy conservation for quantum repeaters under uncertain fidelity requirements. Experimental results demonstrate that our proposed model can reduce the total cost by at least 20\% compared to the baseline model. 
\end{abstract}


\section{Introduction}
\label{sec:introduction}

In recent decades, quantum networks have emerged as a groundbreaking advancement, enabling the support of innovative applications that surpass the capabilities of classical networks. These applications include quantum key distribution (QKD), distributed quantum computing, and quantum encryption protocols~\cite{c-li-effective-routing2021}. Quantum networks rely on entangled qubit pairs, which serve as a fundamental component for end-to-end quantum communication between two quantum nodes. This unique feature allows for secure communication, robust computational power, and novel cryptographic schemes that are resistant to current and future threats. Furthermore, quantum networks pave the way for the development of new technologies and applications that leverage quantum phenomena such as superposition and entanglement, providing substantial advantages over their classical counterparts in terms of speed, security, and efficiency. Consequently, the exploration of quantum networks and their potential continues to gain momentum, driving research into their optimization and integration within existing communication infrastructures.

In quantum networks, quantum nodes are interconnected by optical fiber links~\cite{y-cao-hybrid-trusted-untrusted2021}. Quantum nodes possess the capability to generate quantum information and store it within their quantum memories. Furthermore, they can transmit and receive quantum information between nodes~\cite{j-li-fidelity-guaranteed-entanglement2022, s-shi-concurrent-entanglement2020}. Prior to exchanging information, the quantum network must establish an entanglement connection between the nodes and enable the transmission of quantum information encoded as a quantum bit (qubit) over this connection. Consequently, the quantum source node can utilize entangled pairs in the entanglement connection to transmit information to the quantum destination node. When a quantum source node is distant from the quantum destination node, entanglement connections are generated based on routing, and quantum repeaters (i.e., intermediate quantum nodes in the routing) connect the quantum source node to the quantum destination node using entanglement swapping, or joint Bell state measurements at quantum repeaters, for a remote entanglement connection~\cite{c-li-effective-routing2021}. For large-scale quantum networks, efficiently utilizing entangled pairs and identifying optimal routing for entanglement connections are vital challenges. By optimizing the use of entangled pairs and routing, the energy consumption of quantum repeaters can be minimized in quantum networks.

%\textcolor{red}{Fidelity}
Entanglement fidelity is a crucial factor in guaranteeing the quality of remote entanglement connections, as quantum repeaters may not generate entangled pairs with the desired fidelity due to system noise~\cite{j-li-fidelity-guaranteed-entanglement2022}. Low-fidelity entangled pairs can impact the quality of services provided by quantum applications~\cite{a-s-cacciapuoti-quantum-internet2020}. For instance, the security of key distribution in quantum cryptography protocols, such as the BB84 protocol, can be compromised if entanglement fidelity is lower than the quantum bit error rate requirements~\cite{q-jia-an-improved2021}. Nevertheless, entanglement purification techniques can increase the fidelity value of entangled pairs~\cite{a-s-cacciapuoti-when-entanglement2020}. These techniques use additional entangled pairs to achieve higher fidelity values, but determining the optimal number of additional entangled pairs to satisfy uncertain fidelity requirements for quantum applications remains a challenge and is often overlooked in existing works.

To address these challenges, in this paper, we propose a stochastic resource management framework for achieving optimal entangled resources in quantum networks and introduce a dynamic entanglement purification algorithm to adaptively increase fidelity values. We consider entangled pair resource allocation and fidelity-guaranteed entanglement routing within quantum networks. Specifically, we solve the optimization problem to obtain the optimal number of entangled pairs and fidelity values, satisfying all requests (i.e., multiple quantum source nodes and quantum destination nodes), while considering the uncertainty of fidelity requirements during resource allocation and routing.

The major contributions of this paper can be summarized as follows:
\begin{itemize}
    \item We propose a novel entangled pair resource allocation and fidelity-guaranteed entanglement routing model under uncertainty of fidelity requirements in quantum networks. Additionally, we introduce a dynamic entanglement purification algorithm to elastically improve fidelity values.
    \item We formulate and solve a two-stage stochastic programming (SP) model to obtain not only the optimal decisions on entangled pair resource allocation but also fidelity-guaranteed entanglement routing with the minimum number of quantum repeaters in quantum networks. In the proposed model, the entangled pair resource allocation and fidelity-guaranteed entanglement routing are jointly calculated, with statistical information in the first stage and realization in the second stage.
    \item We evaluate the performance of the proposed model through comprehensive experiments under real-world network topologies. Moreover, we compare the solution of the proposed model with those of baseline models to demonstrate the superior performance of our approach. 
\end{itemize}

\vspace{-0.05cm}
\section{Related Work}
\label{sec:relatedwork}

In this section we provide a brief overview of relevant works in the field, highlighting their contributions and limitations. The authors of~\cite{j-li-fidelity-guaranteed-entanglement2022} presented a fidelity-guaranteed entanglement routing scheme to ensure fidelity for source-destination pairs in quantum networks. In~\cite{j-li-fidelity-guaranteed-entanglement2022}, they initially proposed an iterative routing algorithm (Q-PATH) for optimal entanglement routing with minimum entangled pair cost for single source-destination pairs. For multiple source-destination pairs, they introduced a greedy-based algorithm to minimize the entanglement routing path and entangled pair count. Similarly,\cite{c-li-effective-routing2021} authors proposed an efficient routing scheme for multiple entanglement generation requests in quantum lattice networks with limited quantum resources. Their objective was to allocate quantum resources effectively, meeting entanglement generation requests and fidelity thresholds. In\cite{k-chakraborty-entanglement-dist2020}, the authors suggested a linear programming model to maximize the achievable expected entanglement generation rate between multiple source-destination pairs in a quantum network, satisfying the end-to-end fidelity demand. This problem resembled that in~\cite{j-li-fidelity-guaranteed-entanglement2022}. Nonetheless,\cite{k-chakraborty-entanglement-dist2020} did not consider the purification process. Therefore, in\cite{y-zhao-redundant-entanglement2021}, the authors introduced redundant entanglement provisioning and selection (REPS) for throughput maximization in multi-hop quantum networks with multiple source-destination pairs. The authors of~\cite{l-gyongyosi-adaptive-routing2019} proposed an adaptive routing scheme addressing quantum memory failures in quantum nodes within quantum networks, finding the shortest entanglement paths between source and destination quantum nodes. The author of~\cite{m-caleffi-optimal-routing2017} suggested an optimal routing protocol for quantum networks, identifying the path with the highest end-to-end entanglement rate between source-destination pairs in quantum networks. In~\cite{f-hahn-quantum-network2019}, the authors employed graph-theoretic tools (i.e., graph states) to reduce the number of necessary measurements and proposed a routing method for quantum communication between source and destination nodes. 

However, none of these existing works address the problem of jointly managing entangled pair resources and optimizing fidelity-guaranteed entanglement routing under fidelity requirement uncertainty.

\vspace{-0.05cm}
\section{System Model}
\label{sec:systemmodel-assumptions}

\begin{comment}

%%%%%%%%%%%Entanglement pairs %%%%%%%%%%%%%%
\begin{figure*}[htb]
\centering
\captionsetup{justification=centering}
$\begin{array}{c} \epsfxsize=5.5 in \epsffile{./pics/Entanglement-teleportation-3.eps} \\
\end{array}$
\caption{ Short-distance and long-distance quantum teleportation for qubit transmission in a quantum network.} 
\label{fig:teleportation}
\end{figure*}

\end{comment}

%%%%%%%%%%%Entanglement pairs %%%%%%%%%%%%%%
\begin{figure}[t]
\centering
\captionsetup{justification=centering}
$\begin{array}{c} \epsfxsize=2.8 in \epsffile{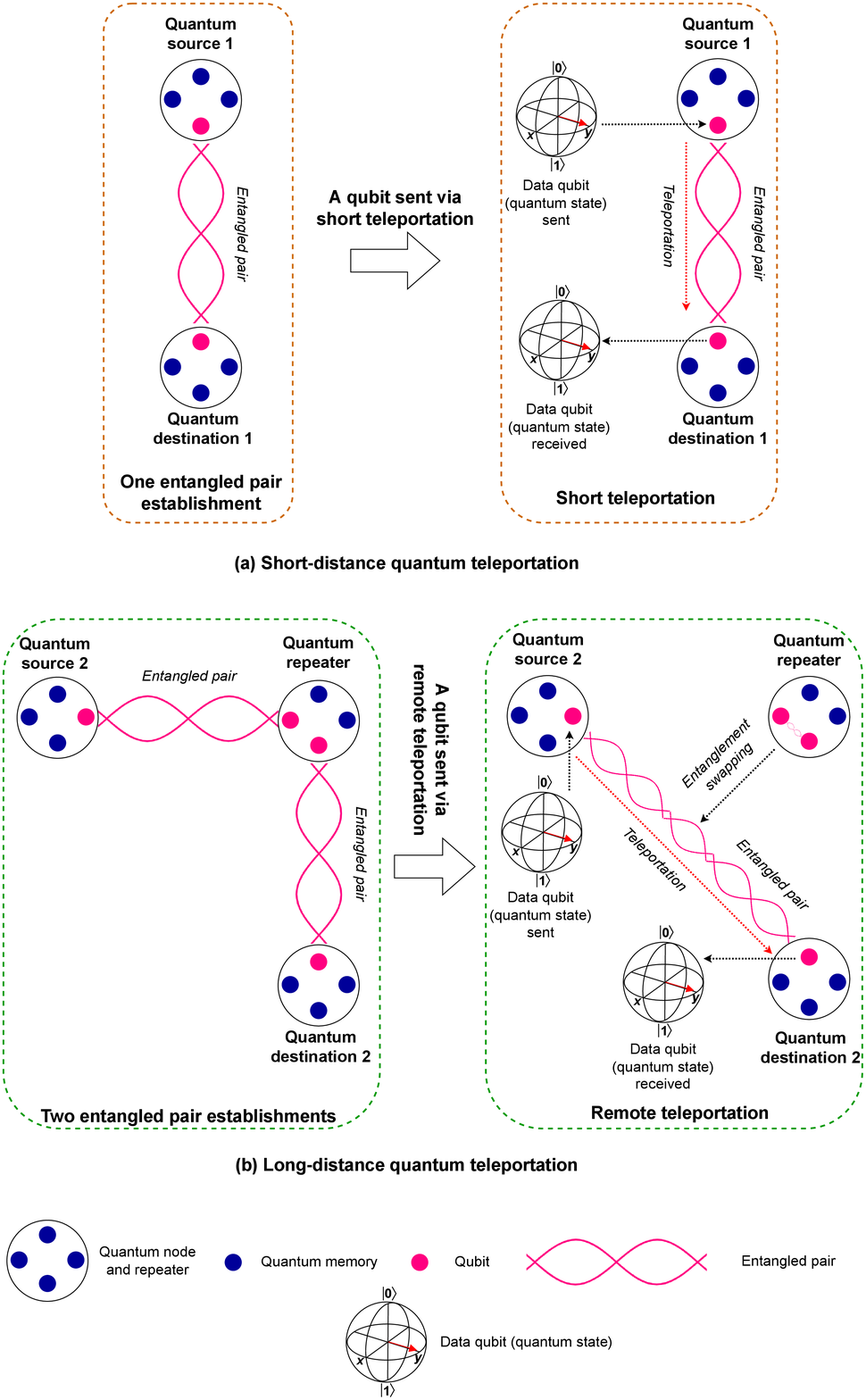} \\
\end{array}$
\caption{ Short-distance and long-distance quantum teleportation for qubit transmission in a quantum network.} 
\label{fig:teleportation}
\vspace{-0.2cm}
\end{figure}

We propose entangled pair resource allocation and fidelity-guaranteed entanglement routing in a quantum network for quantum applications. In the quantum network, quantum nodes connected via optical fiber links can generate, store, exchange, and process quantum information~\cite{j-li-fidelity-guaranteed-entanglement2022,s-shi-concurrent-entanglement2020}, as shown in Figs.~\ref{fig:teleportation}.

In Fig.~\ref{fig:teleportation}, the quantum network establishes an entanglement connection between the source and destination nodes to enable the transmission of quantum information in the form of qubits. To create this connection over long distances, entangled pairs are generated between intermediate quantum nodes located between the source and destination nodes. As illustrated in Fig.~\ref{fig:teleportation}(b), the quantum repeater (an intermediate node) establishes communication with other quantum nodes through entanglement swapping, creating a long-distance entanglement connection. When quantum source node 2 attempts to transmit information to quantum destination node 2, it first becomes entangled with the quantum repeater, which in turn entangles with quantum destination node 2. Subsequently, the repeater performs entanglement swapping, generating a long-distance connection between quantum source node 2 and quantum destination node 2 for the transmission of qubits.

The quantum network is represented by a network graph $G(\mathcal{M},\mathcal{E})$ where $\mathcal{M}$ and $\mathcal{E}$ are a set of quantum nodes and a set of edges between two quantum nodes, respectively. Each quantum node has finite quantum memories that are used to store qubits. The maximum capacity defined as the maximum number of entangled pairs of an edge between quantum nodes $i$ and $j$ is denoted as $C^{\mathrm{etp}}_{i,j}$ where $i$ and $j$ $\in \mathcal{M}$. The entanglement purification is performed on each edge to satisfy the fidelity threshold that is denoted as $F^{\mathrm{ths}}_{i,n}$ where $i$ and $n$ $\in \mathcal{M}$. In each round of the entanglement purification operation, entangled pairs are utilized. The fidelity value of multiple entangled pairs on the same edge is identical while the fidelity value on different edges can vary~\cite{c-li-effective-routing2021}. 

%%%%%%%%%%%Purification process %%%%%%%%%%%%%%
\begin{figure}[t]
\centering
\captionsetup{justification=centering}
$\begin{array}{c} \epsfxsize=3 in \epsffile{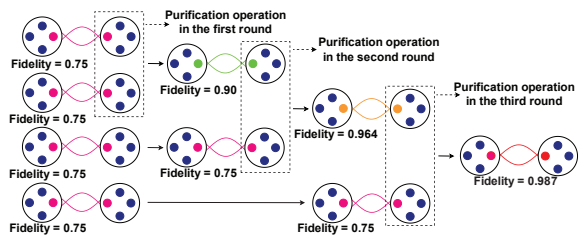} \\
\end{array}$
\caption{An example of three purification rounds in the entanglement purification process.} 
\label{fig:purification-process}
\vspace{-0.3cm}
\end{figure}
\subsection{Network Model}

The quantum node, quantum repeater, quantum source, quantum destination, and quantum channel are described as follows:   

\subsubsection{Quantum Node, Quantum Repeater, Quantum Source, and Quantum Destination} A quantum node can create, exchange, and process quantum information in quantum networks~\cite{s-shi-concurrent-entanglement2020}. The quantum node contains the quantum repeater's function, i.e., entanglement generation, purification, and swapping. Quantum processors and quantum applications can be installed in quantum nodes to establish a quantum network and support the quantum applications. In the quantum repeater, the number of quantum memories is limited, and the entanglement generation, purification, and swapping are applied~\cite{j-li-fidelity-guaranteed-entanglement2022}. In quantum networks, all quantum nodes have limited computing and storage capacities, and are connected via classical networks~\cite{j-li-fidelity-guaranteed-entanglement2022}. A quantum network is typically managed by a centralized controller through classical networks. This controller is responsible for overseeing all the quantum nodes and storing essential information about the network, including its topology and resources. The quantum nodes can report any updates to this information to ensure it remains accurate and up to date. To support the quantum application, a quantum source node can establish an entanglement connection with the quantum destination node according to the requirement of the quantum application.

\subsubsection{Quantum Channel} Each quantum channel established between intermediate quantum nodes is used to share entangled pairs of qubits between the nodes. These entangled pairs are shared via both optical fibers~\cite{j-li-fidelity-guaranteed-entanglement2022, s-shi-concurrent-entanglement2020} and free space\cite{j-g-ren-ground-to-satellite-quantum-teleportation}. Qubits are encoded and then transmitted using quantum teleportation between the entangled pairs of qubits. Therefore, a capacity (i.e., the number of entangled pairs) of a quantum channel between intermediate quantum nodes is generated in advance by the entanglement generation process, e.g., nitrogen-vacancy centers~\cite{p-c-humphreys-deterministic-delivery2018}, before qubit transmission. The entanglement generation process, particularly when using the nitrogen-vacancy center, can be seen as a deterministic black box~\cite{j-li-fidelity-guaranteed-entanglement2022}. In addition, the fidelity of the entangled pair on each quantum channel is approximately computed beforehand by the deterministic equations without noise~\cite{m-caleffi-quantum-switch2020}~\cite{p-c-humphreys-deterministic-delivery2018}, e.g., the deterministic state-delivery protocol~\cite{p-c-humphreys-deterministic-delivery2018}.   

The three steps for an entanglement routing process are described as follows: First, the quantum source node, the quantum destination node, and intermediate quantum nodes generate the entangled pairs in the quantum channel. After that, the network controller calculates the routing and allocates entangled pair resources in the network. Finally, the network controller instructs the corresponding quantum nodes to perform entanglement purification to increase the fidelity of entangled pairs to satisfy the requirement of the quantum application. For the multi-hop entanglement connection, entanglement swapping is introduced to build the long-distance entanglement.  

Entanglement generation, purification, and swapping to establish entanglement connections in the quantum networks can be described as follows: 

\subsubsection{ Entanglement Generation} To establish physical entanglement between two manageable quantum nodes, they are linked to an intermediate station, referred to as the heralding station, through optical fibers. A range of hardware platforms can be employed for this objective, such as nitrogen-vacancy centers in diamond~\cite{a-dahlberg-link-layer2019}. Upon successful generation at the heralding station, the entangled pair is retained in the memory of both quantum nodes. Serving as a valuable resource, the entangled pair facilitates entanglement communication and qubit transmission between the nodes.
%In this paper, we assume that the entangled pairs are created and stored in the memory of quantum nodes in advance. 

\subsubsection{Entanglement Purification} Entanglement purification is applied to increase the fidelity of a Bell pair by combining two low-fidelity Bell pairs into a single high-fidelity Bell pair, which is implemented by controlled-NOT ({\bf C-NOT}) gates or a polarizing beam splitter~\cite{x-m-long-distance-2021}. The entanglement purification function~\cite{j-li-fidelity-guaranteed-entanglement2022} is expressed as follows:
\beqn
    f^{\mathrm{pur}}(q_{1}, q_{2}) = \frac{q_1 q_2}{q_1 q_2 + (1-q_1)(1-q_2)}. \label{entanglement-purification-function}
\eeqn 
$q_1$ and $q_2$ are the fidelity of two Bell pairs in purification operation. The dynamic entanglement purification algorithm is introduced to perform the purification operation in Eq.~(\ref{entanglement-purification-function}) to satisfy the requirement of quantum applications. In the entanglement purification algorithm, each round of purification operation in Eq.~(\ref{entanglement-purification-function}) utilizes an additional entangled pair. For example, Fig.~\ref{fig:purification-process} shows three rounds of entanglement purification operations to increase the fidelity value from 0.75 to 0.987 by utilizing entangled pairs. The entanglement purification algorithm can be expressed in {\bf Algorithm~\ref{algorithm-entanglement-purification}}. The entanglement purification algorithm is applied to the SP model in the constraints of Eqs.~(\ref{eq:def_const7}) and (\ref{eq:def_const8}).   

\subsubsection{Entanglement Swapping} When the quantum source node is far away from the quantum destination node, entanglement swapping is introduced to establish distant entanglement connections along the routing. By using entanglement swapping, the multi-hop entanglement connection can be established along the routing of quantum repeaters containing entangled pairs. 

%%%%%%%%%%%%%%%%%%%%% Algorithm %%%%%%%%%%%%%%%%%%%%%%%%%%%%%%%%%%%
\begin{algorithm}[t]
    \caption{Dynamic Entanglement Purification, i.e., $\mathbf{F}^{\mathrm{epg}}(\cdot)$} \label{algorithm-entanglement-purification}
    \begin{algorithmic}[1]
    \STATE \textbf{Input:} The number of entangled pairs between quantum nodes $i$ and $j$ \\
    \STATE \textbf{Output:} The resulting fidelity value ($f_v$) between quantum nodes $i$ and $j$ \\
%    \STATE \textbf{Output:} The entanglement fidelity value and the number of qubits between node $i$ and node $j$. \\
    \STATE  $N$ is the total number of entangled pairs - 1. \\
    \STATE  $f_v$ is the fidelity value. 
        \FOR{ $p\_round$ = $1$ to $N$}
            \IF {$p\_round$ == 1} 
                  \STATE $q_1$ = the fidelity of the first pair
                  \STATE $q_2$ = the fidelity of the second pair 
                  \STATE $f_v = f^{\mathrm{pur}}(q_1,q_2)$
            \ELSE 
                  \STATE $f_v$ =  $f_v$ of $p\_round - 1$
                  \STATE $q_2$ = the fidelity of the next pair 
                  \STATE  $f_v = f^{\mathrm{pur}}(f_v,q_2)$
            \ENDIF
            \STATE $p\_round = p\_round + 1$
        \ENDFOR
    \end{algorithmic}
\end{algorithm}
%%%%%%%%%%%%%%%%%%%%%% End of algorithm %%%%%%%%%%%%%%%%%%%%%%%%%%%%%%%

\vspace{-0.4cm}
\section{Problem Formulation}
\label{sec:problem-formulation}

\subsection{Model Description}
\label{subsec:model-description}
We define sets and decision variables in the proposed formulation as follows: 
\begin{itemize}
    \item ${\mathcal{M}}$ represents a set of all quantum nodes that are present within the network.
	\item ${\mathcal{Q}}_n$ represents a set of all outbound links from node $n \in {\mathcal{M}}$.
	\item ${\mathcal{J}}_n$ represents a set of all inbound links to node $n \in {\mathcal{M}}$.
	\item ${\mathcal{R}}$ represents a set of requests, i.e., quantum source and destination nodes, in the network.
    \item $x_{i,j,r}$ represents a binary decision variable indicating whether request $r \in \mathcal{R}$ will take a route with the link from nodes $i$ to $j$ or not, i.e., $x_{i,j,r} \in \{0, 1\}$, $i, j \in \mathcal{M}$. 
	\item $y^{\mathrm{r}}_{i,j,r}$ represents a decision variable indicating the number of entangled pairs between nodes $i$ and $j$ in the reservation phase, i.e., $y^{\mathrm{r}}_{i, j, r } \in \{0,1,2,\dots\}$.
	\item $y^{\mathrm{e}}_{i,j,r,\omega}$ represents a decision variable indicating the number of entangled pairs between nodes $i$ and $j$ under scenario $\omega$ in the utilization phase, i.e., $y^{\mathrm{e}}_{i,j,r,\omega} \in \{0,1,2,\dots\}$. 
    \item $y^{\mathrm{o}}_{i,j,r,\omega}$ represents a decision variable indicating the number of entangled pairs between nodes $i$ and $j$ under scenario $\omega$ in the on-demand phase, i.e., $y^{\mathrm{o}}_{i,j,r,\omega} \in \{0,1,2,\dots\}$. 
\end{itemize}
%Note that the number of entangled pairs implies the number of qubits applying for entanglement communication and qubit transmission in the quantum network.

We consider the uncertainty of fidelity requirements for request $r$ in the SP model. As such, fidelity requirements are treated as uncertain parameters. Let $\tilde{\omega}$ and $\omega$ represent the random variables of fidelity requirements and a scenario of request $r$, respectively. A scenario represents a realization of the random variable $\tilde{\omega}$, and its value can be taken from the set of scenarios. We denote the set of all scenarios for each fidelity requirement (i.e., a scenario space), as $\Upsilon$. We denote the set of all scenarios for request $r$ as $\Omega_r$. The set of all scenarios for each fidelity requirement is described as follows:
\beqn
    \Upsilon &=& {\displaystyle \prod_{r \in \mathcal{R}}} \Omega_{r} = \Omega_{1} \times \Omega_{2} \times \dots \times \Omega_{|\mathcal{R}|} \\
    \mbox{Where} & & \Omega_r = \{0.0, \dots ,1.0\}.
\eeqn

Therefore, $\omega$, $\times$, and $|\mathcal{R}|$ are the scenario space of request $r$ (i.e., $\omega \in \Omega_{r}$), the Cartesian product, and the cardinality of the set $\mathcal{R}$, respectively. The probability that the fidelity requirement of request $r$ is realized can be denoted as $\mathbb{P}_{r}(\omega)$.   

\subsection{Stochastic Programming Formulation}
\label{subsec:sp}

We propose the two-stage SP model~\cite{Brige1997} to provision entangled pair resources and fidelity-guaranteed entanglement routing in quantum networks for quantum applications. In the first stage, decisions on provisioning the number of entangled pairs and finding the routing according to the fidelity requirements are performed. In the second stage, when the numbers of entangled pairs in the first stage are inadequate, the number of entangled pairs in the on-demand phase is provisioned to satisfy the rest of the fidelity requirements.

The objective function is designed to achieve the minimum total cost of entangled pair utilization of all quantum nodes to satisfy all the requests, which is expressed as follows: 

%%%%%%%%%%%%%%% Stochastic version %%%%%%%%%%%%%%%%%%%%
\beqn
	\min_{x_{i,n,r}, y^{\mathrm{r}}_{i,n,r}}  & & \sum_{r \in {\mathcal{R}}} \sum_{n \in {\mathcal{M}} } \sum_{i \in {\mathcal{J}}_n } \big( ( E^{\mathrm{eng}}_{n, r } + S^{\mathrm{stp}}_{n, r} )  x_{i,n,r}  \nonumber \\ 
    & & + R^{\mathrm{r}}_{n, r} y^{\mathrm{r}}_{i,n,r} \big) +  {\mathbf{E}} \left[ {\mathscr{L}} (y^{\mathrm{r}}_{i,n,r}, \tilde{\omega} ) \right].  \label{eq:sp_obj} 
\eeqn

Theoretically, we can convert the SP model with the random variable $\tilde{\omega}$ into a deterministic equivalent formulation~\cite{Brige1997}, expressed in Eqs.~(\ref{eq:def_obj}) - (\ref{eq:def_const9}). The objective function in Eq. (\ref{eq:def_obj}) corresponds to and shares the same meaning as Eq. (\ref{eq:sp_obj}). Eq. (\ref{eq:def_const1}) enforces that the number of outbound routes exceeds the number of inbound routes if the node is the source node $S_r$ of request $r$. Eq. (\ref{eq:def_const2}) dictates that the number of inbound routes surpasses the number of outbound routes if the node is the destination node $D_r$ of request $r$. Eq. (\ref{eq:def_const3}) requires that the number of outbound routes equals the number of inbound routes if the node serves as an intermediate node for request $r$. Eq. (\ref{eq:def_const4}) ensures no loop for any request, implying that each node has only one outbound route for the request. Eq. (\ref{eq:def_const5}) establishes that the number of reserved entangled pairs between node $i$ and node $n$ in the reservation phase does not exceed the maximum capacity of entangled pairs between node $i$ and node $j$ ($C^{\mathrm{etp}}_{i,j}$). Eq. (\ref{eq:def_const6}) asserts that the number of utilized entangled pairs between node $i$ and node $n$ in the utilization phase is not greater than the number of reserved entangled pairs between node $i$ and node $n$ in the reservation phase. Eq. (\ref{eq:def_const7}) states that the numbers of entangled pairs in utilization and on-demand phases must meet the entanglement fidelity requirement. $\mathbf{F}^{\mathrm{epg}}(\cdot)$ in Eq. (\ref{eq:def_const7}) refers to the entanglement purification algorithm applied to calculate entanglement fidelity based on the numbers of entangled pairs in utilization and on-demand phases. Eq. (\ref{eq:def_const8}) stipulates that the numbers of entangled pairs in utilization and on-demand phases must satisfy the entanglement fidelity threshold. Finally, Eq. (\ref{eq:def_const9}) ensures that the number of entangled pairs used in the on-demand phase does not surpass the maximum capacity of entangled pairs between node $i$ and node $j$ ($O^{\mathrm{etp}}_{i,j}$).

%The constraint in Eq. (\ref{eq:def_const10}) defines that the decision variables are binary variables. The constraint in Eq. (\ref{eq:def_const11}) defines that the decision variables are non-negative integers.

%%%%%%%%%%%%%%% deterministic equivalent formulation version %%%%%%%%%%%%%%%%%%%%
%\begin{figure*}[htb]
\vspace{-0.4cm}
\beqn
	& & \min_{x_{i,n,r}, y^{\mathrm{r}}_{i,n,r}, y^{\mathrm{e}}_{i,n,r,\omega}, y^{\mathrm{o}}_{i,n,r,\omega}} \sum_{r \in {\mathcal{R}}} \sum_{n \in {\mathcal{M}} } \sum_{i \in {\mathcal{J}}_n } \label{eq:def_obj}  \nonumber \\
    & &	\big( ( E^{\mathrm{eng}}_{n} + S^{\mathrm{stp}}_{n} ) x_{i,n,r} y^{\mathrm{r}}_{i,n,r} + R^{\mathrm{r}}_{n,r} y^{\mathrm{r}}_{i,n,r} \big) + \sum_{r \in {\mathcal{R}}} \nonumber \\
	& &  \Big( \mathbb{P}_{r}(\omega) \sum_{n \in {\mathcal{M}} } \sum_{i \in {\mathcal{J}}_n } \big( U^{\mathrm{e}}_{n, r} y^{\mathrm{e}}_{i,n,r,\omega} + O^{\mathrm{o}}_{n, r} y^{\mathrm{o}}_{i,n,r,\omega}  \big) \Big)  \\
\mbox{s.t.} 
	%%%%%% routing constraints %%%%%%		
	& & \sum_{ j' \in {\mathcal{Q}}_{S_r} }	x_{S_r,j',r}	-	\sum_{ i' \in {\mathcal{J}}_{S_r} }	x_{i',S_r, r} =	1,	 r	\in {\mathcal{R}},	\label{eq:def_const1} \\
	& & \sum_{ i' \in {\mathcal{J}}_{D_r} } x_{i', D_r, r} - \sum_{ j' \in {\mathcal{Q}}_{D_r} } x_{D_r, j', r } =	1,	 r \in {\mathcal{R}},	 \label{eq:def_const2} \\
	& & \sum_{ j' \in {\mathcal{Q}}_n } x_{n, j', r }	-	\sum_{i' \in {\mathcal{J}}_n } x_{i',n, r}	=	0,	 r	\in {\mathcal{R}}, \nonumber \\
	& &  n \in {\mathcal{M}} \setminus \{ S_r, D_r \}, \label{eq:def_const3} \\
	& & \sum_{j' \in {\mathcal{Q}}_n } x_{n, j', r } 	\leq	1,	 n \in {\mathcal{M}}, r \in {\mathcal{R}}, \label{eq:def_const4} \\ 
	%%%%%% expending and on-demand phases %%%%%%
    & & \sum_{r \in {\mathcal{R}}}  y^{\mathrm{r}}_{i,n,r} x_{i,n, r} \leq C^{\mathrm{etp}}_{i,j}, i,j,n \in {\mathcal{M}}, \label{eq:def_const5} \\ 
	& & y^{\mathrm{e}}_{i,n,r,\omega} x_{i, n, r} \leq y^{\mathrm{r}}_{i,n,r} x_{i,n,r},  \nonumber \\ 
    & & i,j,n \in {\mathcal{M}}, r \in {\mathcal{R}}, \forall \omega \in \Omega_{r}, \label{eq:def_const6} \\
	& & \mathbf{F}^{\mathrm{epg}}\big( ( y^{\mathrm{e}}_{i,n,r,\omega} x_{i,j, r}) + y^{\mathrm{o}}_{i,n,r,\omega} \big) \geq \omega, \nonumber \\ 
	& & i,n \in {\mathcal{M}}, r \in {\mathcal{R}}, \omega \in \Omega_{r}, \label{eq:def_const7}\\  
    & & \mathbf{F}^{\mathrm{epg}}\big( ( y^{\mathrm{e}}_{i,n,r,\omega} x_{i,j,r}) + y^{\mathrm{o}}_{i,n,r,\omega} \big) \geq F^{\mathrm{ths}}_{i,n}, \nonumber \\ 
	& & i,n \in {\mathcal{M}}, r \in {\mathcal{R}}, \omega \in \Omega_{r}, \label{eq:def_const8} \\ 
    & & \sum_{r \in {\mathcal{R}}} \big( y^{\mathrm{o}}_{i,j,r, \omega}  x_{i,j, r} \big) \leq O^{\mathrm{etp}}_{i,j}, i,j \in {\mathcal{M}}, \forall \omega \in \Omega_{r}. \label{eq:def_const9} 
%	& & x_{i,n,f} \in \{ 0, 1 \}, i,n \in {\mathcal{M}}, f \in {\mathcal{R}},  \label{eq:def_const10} \\
%	& & y^{\mathrm{r}}_{i, n, f }, y^{\mathrm{e}}_{i, n, f, \omega }, y^{\mathrm{o}}_{i, n, f, \omega } \in \{0,1,2,\dots\},  \nonumber \\
%	& & i,n \in {\mathcal{M}}, f \in {\mathcal{R}}, \forall \omega \in \Omega_{f}.  \label{eq:def_const11}  	
\eeqn
%The number of entangled pairs implies the number of qubits that are used for the entanglement establishment

\section{Performance Evaluation}
\label{sec:performanceEvaluation}

\subsection{Parameter Setting}

\begin{figure*}[htb]
 \centering
 \captionsetup{justification=centering}
 \subfloat[Three requests.]{\label{fig:routing-three-requests}\includegraphics[width=0.25\textwidth]{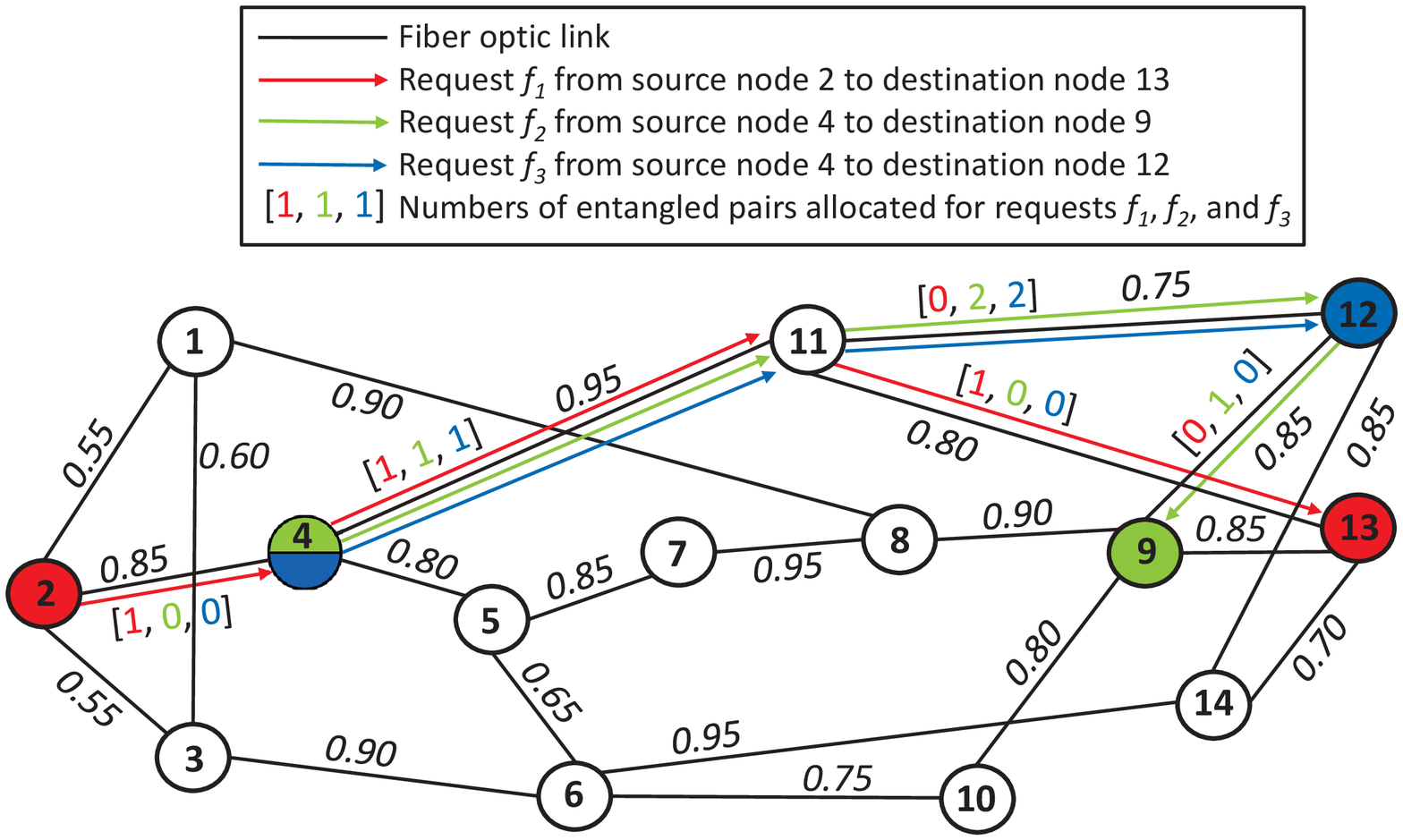}}
 \subfloat[The entangled pair utilization.]{\label{fig:reserved-utilized-on-demand-entangled-pairs}\includegraphics[width=0.25\textwidth]{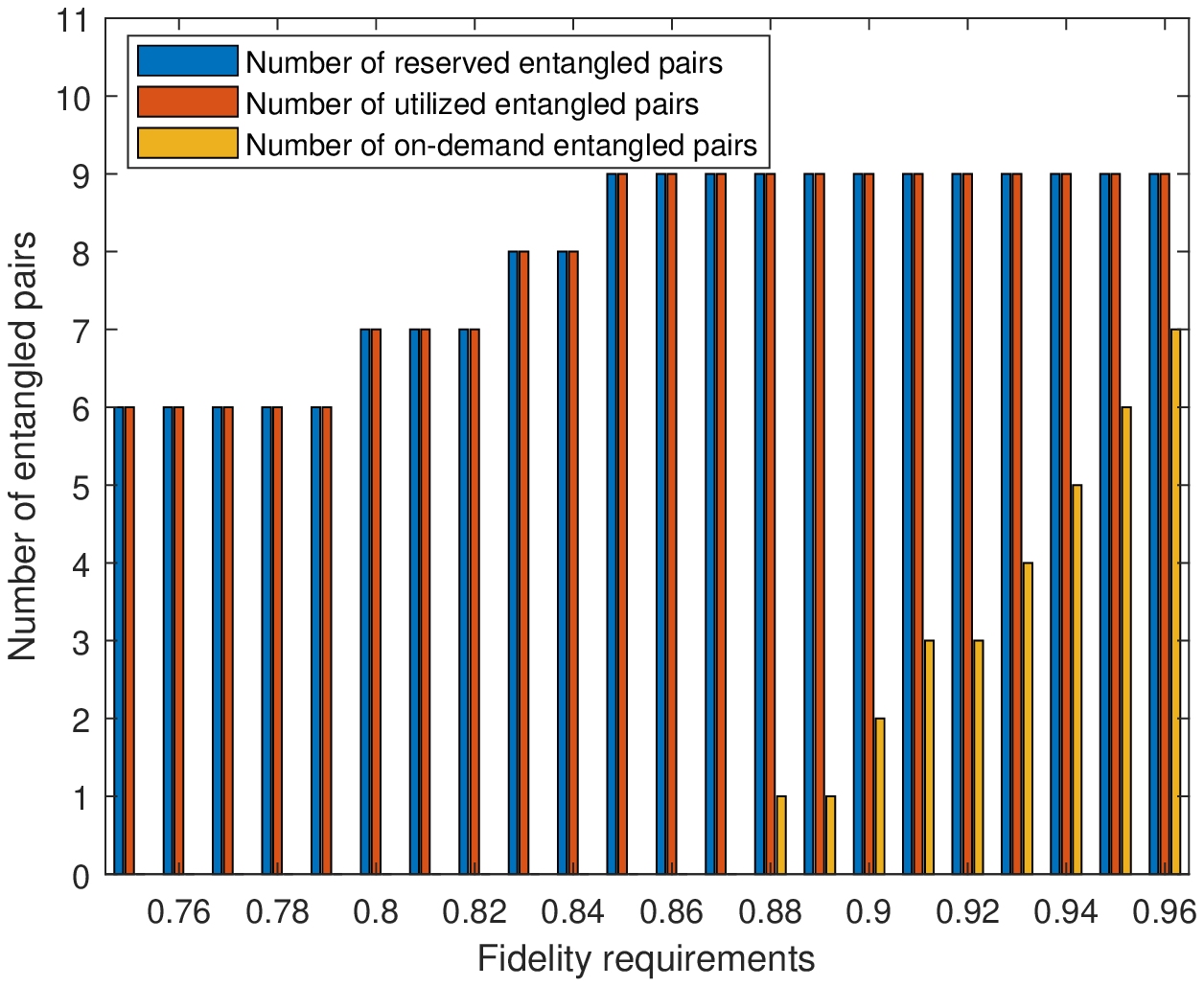}}
 \subfloat[The optimal solution.]{\label{fig:optimal-solution-different-entangled-pairs}\includegraphics[width=0.25\textwidth]{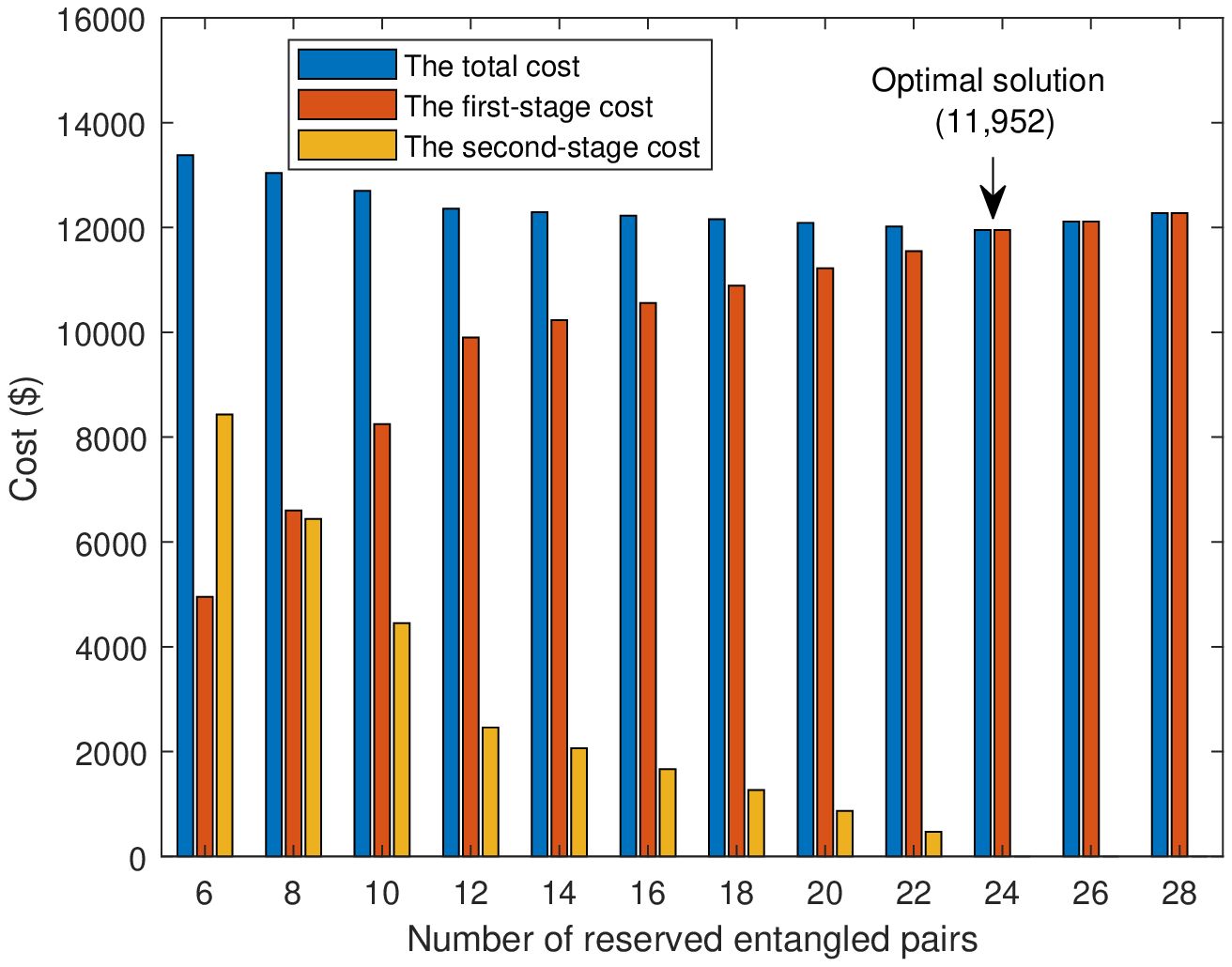}}
 \subfloat[The cost comparison.]{\label{fig:three-model-comparison}\includegraphics[width=0.25 \textwidth]{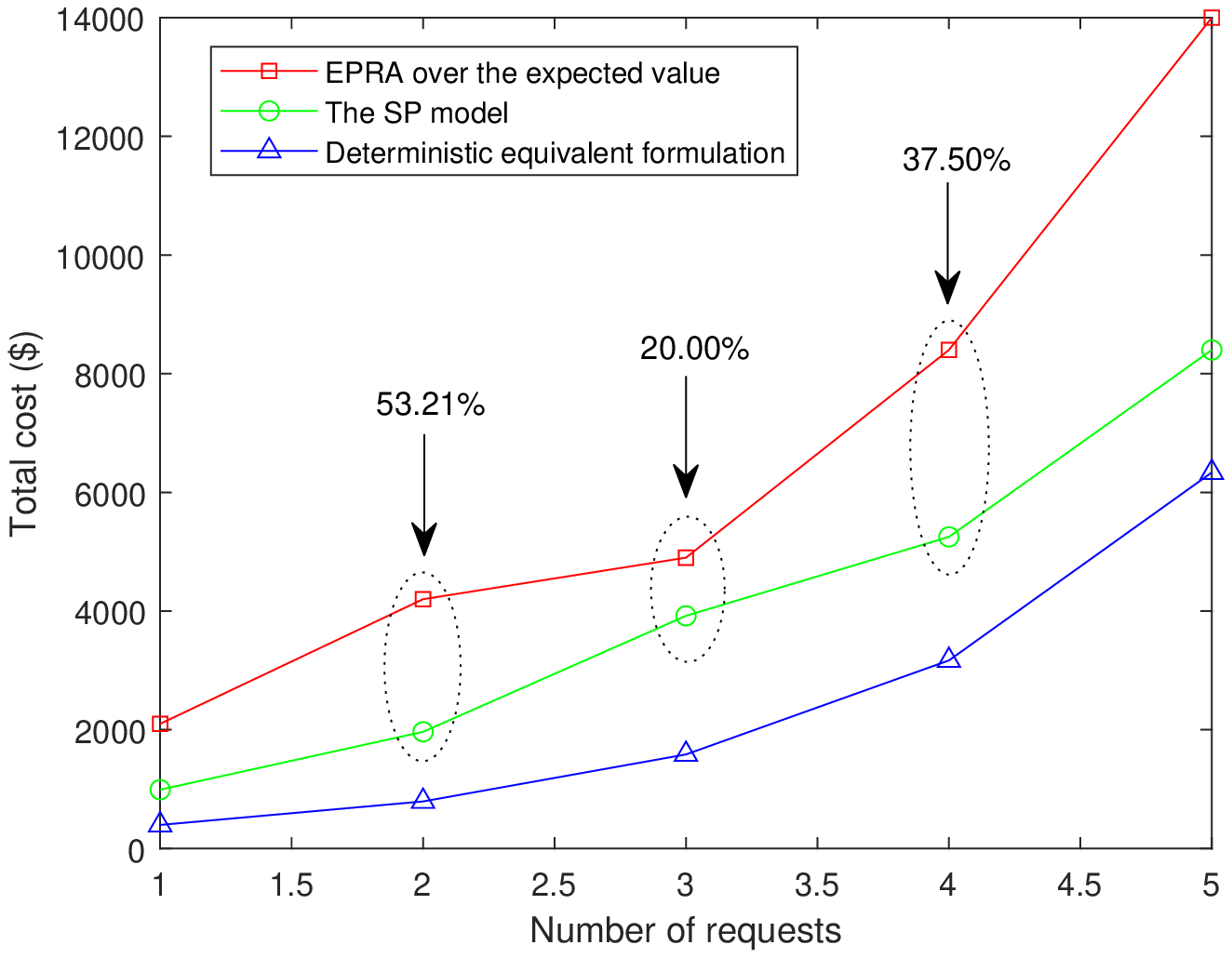}}
 \caption{(a) Three requests in NSFNET topology, (b) The entangled pair utilization under different fidelity requirements, (c) The optimal solution under different numbers of reserved entangled pairs, and (d) The cost comparison of three models.}
 \label{fig:entangled-pair-and-optimal-solutions}
 \vspace{-0.3cm}
\end{figure*}

We examine the network topology of NSFNET, which is connected using optical fibers~\cite{y-cao-hybrid-trusted-untrusted2021} and perform experiments on this topology. For each quantum node in the topology, we set the fidelity values between nodes $i$ and $j$ as shown in Fig.~\ref{fig:entangled-pair-and-optimal-solutions}(a). The fidelity threshold is initially 0.8~\cite{j-li-fidelity-guaranteed-entanglement2022}. The maximum numbers of entangled pairs between nodes $i$ and $j$ in the reservation phase ($C^{\mathrm{etp}}_{i,j}$) and the on-demand phase ($O^{\mathrm{etp}}_{i,j}$ ) are initially 10 and 60, respectively. For the SP model, we consider a random number of requests with fidelity requirements that are uniformly distributed. We assume the costs of reservation, utilization, and on-demand phases are 10\$, 1\$, and 200\$, respectively. The cost of energy consumption of transferring traffic of request $r$ through node $n$ ($E^{\mathrm{eng}}_{n}$) is 5\$. The cost of energy consumption to establish repeater $n$ ($S^{\mathrm{stp}}_{n}$) is 150\$. We implement and solve the entangled pair resource allocation and fidelity-guaranteed entanglement routing via the GAMS/CPLEX solver~\cite{Gams}.

\subsection{Numerical Results}

%%%%%%%%%%%%%%%%%%%%%%%%%%%%%%%%%%%%%%%%%%%%%%%%%%%%%%%%%%%%%%%%%%%%%%%%%
%%%%%%%%%%%%%%%%%%%%%%%%%%%%%%%%%%%%%%%%%%%%%%%%%%%%%%%%%%%%%%%%%%%%%%%%%

\subsubsection{Routing and Entangled Pair Utilization}

Figure~\ref{fig:entangled-pair-and-optimal-solutions}(a) shows the solutions of the SP model that satisfies the fidelity requirements of three requests (i.e., $f_1$, $f_2$, and $f_3$) represented by different colors. In the solutions, the SP model not only allocates the entangled pairs to satisfy fidelity requirements but also minimizes the number of repeaters (i.e., intermediate quantum nodes) for the requests in the network. In Fig.~\ref{fig:entangled-pair-and-optimal-solutions}(a), to obtain the optimal cost, the entangled pairs are forced to share the same edges and the number of repeaters is forcibly minimized while supporting all requests. For example, each request utilizes one entangled pair (i.e., $[1,1,1]$) on the edge between quantum nodes 4 and 11, which is a direct result of minimizing the number of repeaters. 

Figure~\ref{fig:entangled-pair-and-optimal-solutions}(b) shows the number of entangled pairs in the reservation, utilization, and on-demand phases under different fidelity requirements. In Fig.~\ref{fig:entangled-pair-and-optimal-solutions}(b), in the reservation and utilization phases, the number of entangled pairs reserved and then utilized increases steadily until the fidelity requirement is 0.87. At this point, the numbers of reserved and utilized entangled pairs reach the maximum capacity of the entangled pairs. As a result, the numbers of reserved and utilized entangled pairs are stable at 9 entangled pairs, and it cannot support the higher fidelity requirements. Therefore, to meet the high fidelity requirements, the entangled pairs in the on-demand phase are utilized. In the on-demand phase, the number of entangled pairs starts utilizing at 0.88 of the fidelity requirement due to the limited capacity of the entangled pairs in the reservation phase.

\subsubsection{Cost Structure Analysis}

In Fig.~\ref{fig:entangled-pair-and-optimal-solutions}(c), we assess the SP model's effectiveness in attaining an optimal solution. Initially, we modify the number of reserved entangled pairs and subsequently present the optimal solution derived from the SP model, as well as the impact of reserved entangled pairs on this solution. In Fig.~\ref{fig:entangled-pair-and-optimal-solutions}(c), the first-stage cost notably escalates as the quantity of reserved entangled pairs expands. In contrast, the second-stage cost markedly diminishes when fidelity requirements are satisfied. This is a consequence of the reservation phase (i.e., the first stage) necessitating the maximum number of entangled pairs due to lower costs, while the on-demand phase (i.e., the second stage) demands the minimum number of entangled pairs. As a result, with 24 reserved entangled pairs, the optimal solution is attained at 11,952\$, and the second-stage cost remains at 0. This occurs because the reserved entangled pairs fulfill fidelity requirements, negating the need for on-demand entangled pairs in the second stage. Beyond 24 reserved entangled pairs, the total cost and first-stage cost experience a slight increase due to a penalty cost for surplus reserved entangled pairs. In Fig.~\ref{fig:entangled-pair-and-optimal-solutions}(c), we observe that over- and under-provision of entangled pairs contribute to the overall high total cost.

\subsubsection{Performance Evaluation}

We compare the SP model's performance with two alternative models: the entangled pair resource allocation (EPRA) over the expected value and the deterministic equivalent formulation. In the EPRA over the expected value, the fidelity requirements in the first stage are treated as expected demands. In the deterministic equivalent formulation, fidelity requirements are considered as exact demands. In Fig.~\ref{fig:entangled-pair-and-optimal-solutions}(d), the SP model evidently attains the optimal solution in comparison to the EPRA over the expected value as the number of requests increases. For instance, when the request count is 2, the SP model can reduce the total cost by 53.21\% compared to the EPRA over the expected value. However, the SP model's solution is inferior to that of the deterministic equivalent formulation. This difference arises because the deterministic equivalent formulation employs exact fidelity requirements to reach the solution, while the SP model relies on statistical information of fidelity requirements. Nevertheless, the SP model is more pragmatic than the deterministic equivalent formulation, as determining exact entangled pair fidelity requirements for input in the deterministic equivalent formulation is challenging in real-world scenarios.
\vspace{-0.15cm}
\section{Conclusion}
\label{sec:conclusion}

In this paper, we have proposed the entangled pair resource allocation and fidelity-guaranteed entanglement routing model for quantum networks. Using the two-stage SP framework, we have formulated the model to determine the optimal cost under fidelity requirement uncertainty. The experimental results have demonstrated that our proposed model not only achieves optimal total cost and entanglement routing but also minimizes the number of repeaters (i.e., intermediate quantum nodes). Moreover, the model's performance has surpassed that of the EPRA over the expected value by a minimum of 20\%. In future research, we plan to investigate and incorporate the energy consumption model of quantum nodes within quantum networks into the SP model. Additionally, we aim to develop an entangled pair resource allocation and fidelity-guaranteed entanglement routing model for space-air-ground integrated networks (SAGIN).

\begin{comment}
\section*{Acknowledgment}
This research is supported by the National Research Foundation Singapore and DSO National Laboratories under the AI Singapore Programme (AISG Award No: AISG2-RP-2020-019); the National Research Foundation (NRF), Singapore and Infocomm Media Development Authority under the Future Communications Research Development Programme (FCP); Energy Research Test-Bed and Industry Partnership Funding Initiative, part of the Energy Grid (EG) 2.0 programme; DesCartes and the Campus for Research Excellence and Technological Enterprise (CREATE) programme; Alibaba Group through Alibaba Innovative Research (AIR) Program and Alibaba-NTU Singapore Joint Research Institute (JRI); and Nanyang Technological University, Nanyang Assistant Professorship.
\end{comment}


\vspace{-0.25cm}
\begin{thebibliography}{00}

\bibitem{c-li-effective-routing2021}
C. Li {\em et al.}, ``Effective routing design for remote entanglement generation on quantum networks,'' {\em NPJ Quantum Inf.}, vol. 7, no. 1, pp. 1-12, 2021.

%Hybrid trusted/untrusted QKD 
\bibitem{y-cao-hybrid-trusted-untrusted2021}
Y. Cao {\em et al.}, ``Hybrid Trusted/Untrusted Relay-Based Quantum Key Distribution Over Optical Backbone Networks,'' {\em IEEE JSAC}, vol. 39, no. 9, pp. 2701-2718, 2021.

\bibitem{j-li-fidelity-guaranteed-entanglement2022}
J. Li {\em et al.}, ``Fidelity-Guaranteed Entanglement Routing in Quantum Networks,'' in {\em IEEE Transactions on Communications}, vol. 70, no. 10, pp. 6748-6763, 2022.

\bibitem{s-shi-concurrent-entanglement2020}
S. Shi and C. Qian, ``Concurrent entanglement routing for quantum networks: Model and designs,'' in {\em Proc. ACM SIGCOMM Conf.}, pp. 62–75, 2020.

\bibitem{a-s-cacciapuoti-quantum-internet2020} 
A. S. Cacciapuoti {\em et al.}, ``Quantum Internet: Networking Challenges in Distributed Quantum Computing,'' in {\em IEEE Network}, vol. 34, no. 1, pp. 137-143, 2020.
%A. S. Cacciapuoti, M. Caleffi, F. Tafuri, F. S. Cataliotti, S. Gherardini, and G. Bianchi, ``Quantum Internet: Networking Challenges in Distributed Quantum Computing,'' in {\em IEEE Network}, vol. 34, no. 1, pp. 137-143, 2020.

\bibitem{q-jia-an-improved2021}
Q. Jia {\em et al.}, ``An improved QKD protocol without public announcement basis using periodically derived basis,'' {\em Quantum Inf. Process}, vol. 20, 69 (2021).

\bibitem{a-s-cacciapuoti-when-entanglement2020}  
A. S. Cacciapuoti {\em et al.}, ``When Entanglement Meets Classical Communications: Quantum Teleportation for the Quantum Internet,'' in {\em IEEE Transactions on Communications}, vol. 68, no. 6, pp. 3808-3833, 2020.
%A. S. Cacciapuoti, M. Caleffi, R. Van Meter, and L. Hanzo, ``When Entanglement Meets Classical Communications: Quantum Teleportation for the Quantum Internet,'' in {\em IEEE Transactions on Communications}, vol. 68, no. 6, pp. 3808-3833, 2020.

\bibitem{k-chakraborty-entanglement-dist2020}
K. Chakraborty {\em et al.}, ``Entanglement Distribution in a Quantum Network: A Multicommodity Flow-Based Approach,'' in {\em IEEE TQE}, vol. 1, pp. 1-21, 2020.

\bibitem{y-zhao-redundant-entanglement2021}
Y. Zhao and C. Qiao, ``Redundant Entanglement Provisioning and Selection for Throughput Maximization in Quantum Networks,'' {\em IEEE INFOCOM 2021}, pp. 1-10, 2021.

\bibitem{l-gyongyosi-adaptive-routing2019}
L. Gyongyosi and S. Imre, ``Adaptive routing for quantum memory failures in the quantum internet,'' {\em Quantum Inf. Process}, vol. 18, no. 2, pp. 1–21, 2019.

\bibitem{m-caleffi-optimal-routing2017}
M. Caleffi, ``Optimal Routing for Quantum Networks,'' {\em IEEE Access}, vol. 5, pp. 22299-22312, 2017.

\bibitem{f-hahn-quantum-network2019}
F. Hahn, A. Pappa, and J. Eisert, ``Quantum network routing and local complementation,'' {\em  NPJ Quantum Inf.}, vol. 5, no. 1, pp. 1–7, 2019.

\bibitem{j-g-ren-ground-to-satellite-quantum-teleportation}
J.-G. Ren {\em et al.}, ``Ground-to-satellite quantum teleportation,'' {\em Nature}, vol. 549, no. 7670, pp. 70–73, 2017.

\bibitem{m-caleffi-quantum-switch2020}
M. Caleffi and A. S. Cacciapuoti, ``Quantum Switch for the Quantum Internet: Noiseless Communications Through Noisy Channels,'' in {\em IEEE JSAC}, vol. 38, no. 3, pp. 575-588, 2020.

\bibitem{p-c-humphreys-deterministic-delivery2018}
P. C. Humphreys {\em et al.}. "Deterministic delivery of remote entanglement on a quantum network." Nature, vol. 558, no. 7709, pp. 268-286, 2018.

\bibitem{a-dahlberg-link-layer2019}
A. Dahlberg {\em et al.}, ``A link layer protocol for quantum networks,'' in {\em Proc. ACM SIGCOMM Conf.}, pp. 159–173, 2019.

\bibitem{x-m-long-distance-2021}
X.-M. Hu {\em et al.}, ``Long-distance entanglement purification for quantum communication,'' {\em Phys. Rev. Lett.}, vol. 126, no. 1, 2021.

%\bibitem{j-k-lenstra-complexity-of-vr1981}
%J. K. Lenstra and A. H. G. Rinnooy Kan, ``Complexity of vehicle routing and scheduling problems'', {\em Networks}, vol. 11, no. 2, pp. 221–227, 1981.

%energy consumption of node n 
%\bibitem{r-kaewpuang-optimal-decentralized2014}
%R. Kaewpuang, D. Niyato, P. Wang, Z. Han, and R. Lu, ``Optimal decentralized security software deployment in multihop wireless networks,'' {\em IEEE Global Communications Conference}, pp. 619-624, 2014.

%stochastic programming
\bibitem{Brige1997}
J. R. Birge, and F. Louveaux, ``Introduction to Stochastic Programming,'' 2nd ed. {\em Springer}, 2011.

%GAMS
\bibitem{Gams}
General Algebraic Modeling System (GAMS), https://www.gams.com/, 2022.

\end{thebibliography}
\end{document}